\documentclass[aps,prl,twocolumn,showpacs]{revtex4}

\usepackage{graphicx}
\usepackage{amsmath,amsfonts,amssymb}		
\usepackage{mathrsfs,bbm}

\def\ketbra#1#2{|#1\rangle\langle#2|}

\begin{document}

\title{Continuous Measurement of Spin Systems with Spatially-Distinguishable Particles}
\author{Ben Q.~Baragiola}
\email{quinn.phys@gmail.com}
\author{Bradley A. Chase}
\author{JM Geremia}
\email{jgeremia@unm.edu}
\affiliation{Department of Physics \& Astronomy, The University of New Mexico, Albuquerque, New Mexico USA}

\begin{abstract}
It is generally believed that dispersive polarimetric detection of  collective angular momentum in large atomic spin systems gives rise to: squeezing in the measured observable, anti-squeezing in a conjugate observable, and collective spin eigenstates in the long-time limit (provided that decoherence is suitably controlled).  We show that such behavior only holds when the particles in the ensemble cannot be spatially distinguished--- even in principle--- regardless of whether the measurement is only sensitive to collective observables.  While measuring a cloud of spatially-distinguishable spin-1/2 particles does reduce the uncertainty in the measured spin component, it generates neither squeezing nor anti-squeezing.   The steady state of the measurement is highly mixed, albeit with a well-defined value of the measured collective angular momentum observable.  
\end{abstract}
\date{\today}



\maketitle

\noindent 

\begin{figure}[t!]
\begin{center} \hspace{4mm} \includegraphics{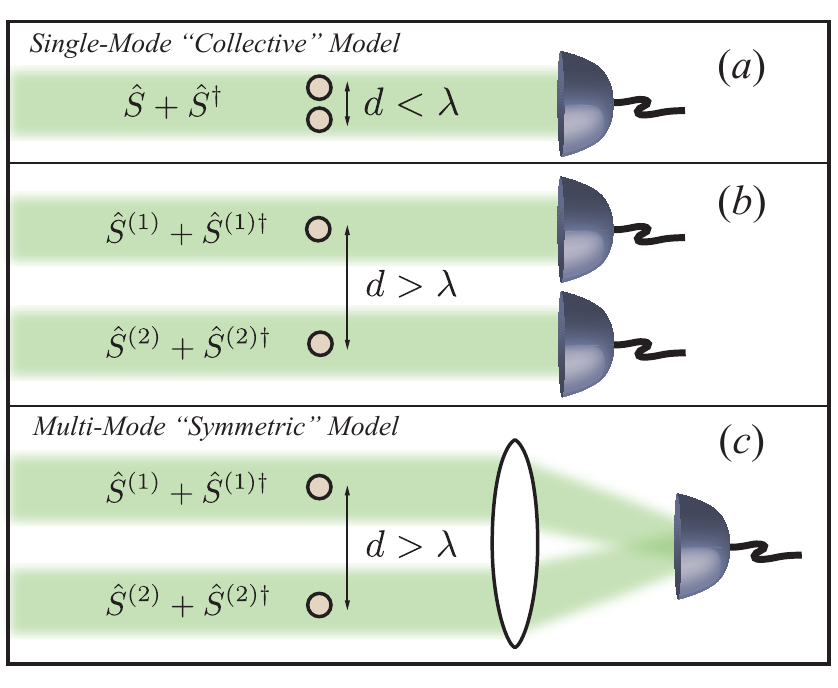} \end{center}
\vspace{-6mm}
\caption{The interaction of a spin ensemble with a free-space probe laser can be analyzed by considering several limiting cases (for clarity only two atoms are depicted here):  $(a)$ if the interparticle distance $d$ is small compared to the optical wavelength $\lambda$, then the particles cannot be distinguished by imaging, and a single-mode field approximation is justified;  $(b)$ if however the interparticle distance $d$ is large compared to the wavelength $\lambda$, then the individual particles can be imaged, and there is a suitable mode decomposition where each atom couples essentially to its own separately-detectable mode.  Typical free-space experiments lie closer to $(b)$ than $(a)$, but attempt to recover a collective spin measurement of the type in $(a)$ by focusing all the scattered spatial modes onto a single detector, as depicted in case $(c)$.  This approach, however, fails to keep track of the ``which-mode'' information that is available under $(b)$ and thus introduces a decoherence mechanism that is ignored in previous treatments \cite{Bouten:2007b,Stockton2004,Takano:2008a}.  \label{Fig:Models}}
\end{figure}

Continuous measurement of hyperfine spin is typically performed by interacting an atomic sample with a probe laser and then detecting the forward-scattered field \cite{Appel2009,Takano:2008a,Windpassinger:2008a,Fernholtz:2008a,Bouten:2007b,Stockton2004}.  In conventional treatments of free-space coupling, the atoms are described by their collective spin \cite{Appel2009,Takano:2008a,Kuzmich:1999a,Takahashi:1999a,Windpassinger:2008a,Fernholtz:2008a,Bouten:2007b,Stockton2004} and the field is approximated explicitly as a single spatial mode \cite{Kuzmich:1999a,Takahashi:1999a,Takano:2008a,vanHandel:2005a,Bouten:2007b}.     However, the size of the atomic sample in typical experiments is large compared to the laser wavelength, thus it is possible (at least in principle) to image the scattered laser field, making a single-mode approximation inadequate \cite{Chase:2009d,footnote1}.  The purpose of this letter is to illustrate that even if the forward-scattered field is focused onto a single detector, models based on a collective, single-mode approximation incorrectly predict even qualitative features of the measurement (please see Fig.~\ref{Fig:Models}). 

\vspace{1mm}


\noindent\textit{Physical Model---} Consider a system of $N$ spin-1/2 particles interacting with a linearly-polarized, far-detuned probe laser propagating along the laboratory $z$-axis.  After adiabatic elimination of excited atomic states, the conditions of Fig.\ \ref{Fig:Models}(a) result in the Faraday-effect interaction Hamiltonian $\hat{H}_C =  \hbar k \hat{J}_\mathrm{z} \hat{s}_{\mathrm{z},t}$ \cite{vanHandel:2005a,Bouten:2007b}, where $\hat{s}_{\mathrm{z},t}$ is the $z$-component of the time-dependent Stokes operator for the single-mode field and the $\hat{J}_i = \sum_{n=1}^N \hat{j}_i^{(n)}$ are collective atomic spin operators with $\hat{j}_i^{(n)} = \hat{\sigma}_i^{(n)}/2$.   Under the conditions of Fig.\ \ref{Fig:Models}(b-c), however, adiabatic elimination of atomic excited states yields an interaction Hamiltonian of the form $\hat{H}_S = \hbar k  \sum_{n=1}^N \hat{j}_\mathrm{z}^{(n)} \hat{s}^{(n)}_{\mathrm{z},t}$, where $\hat{s}_{i,t}^{(n)}$ is the Stokes operator of the field mode that interacts only with atom $n$  \cite{Chase:2009d}.  For simplicity, we assume that the atom-field coupling $k$ is identical for all atoms.

The dynamics of any operator $\hat{O}_\mathrm{AF} = \hat{X}_\mathrm{A} \otimes \hat{Y}_\mathrm{F}$ acting on both the atoms and the field is given by the quantum flow  $j_t[\hat{O}_\mathrm{AF}] \equiv \hat{U}_t^\dagger \hat{O}_\mathrm{AF} \hat{U}_t$, where $\hat{U}_t$ is the propagator for the joint system and can be determined from $\hat{H}_C$ or $\hat{H}_S$ using well-established procedures from quantum stochastic calculus \cite{Hudson:1984a,Bouten:2007b,Chase:2009d}.  Continuous detection of the scattered laser field provides a measurement record $\mathcal{Y}_t$ comprised of the sequence of random measurement outcomes during $0 \le t$.  The conditional expectation value of any atomic operator $\hat{X}$ given the measurement data
\begin{equation}
	\pi_t[\hat{X}] \equiv \mathbbm{E}[\hat{X}|\mathcal{Y}_t]
\end{equation}
can be determined using the methods of quantum filtering theory \cite{vanHandel:2005a,Bouten:2007b,Belavkin:1999a}.


\begin{figure*}[t]
\begin{center} \includegraphics{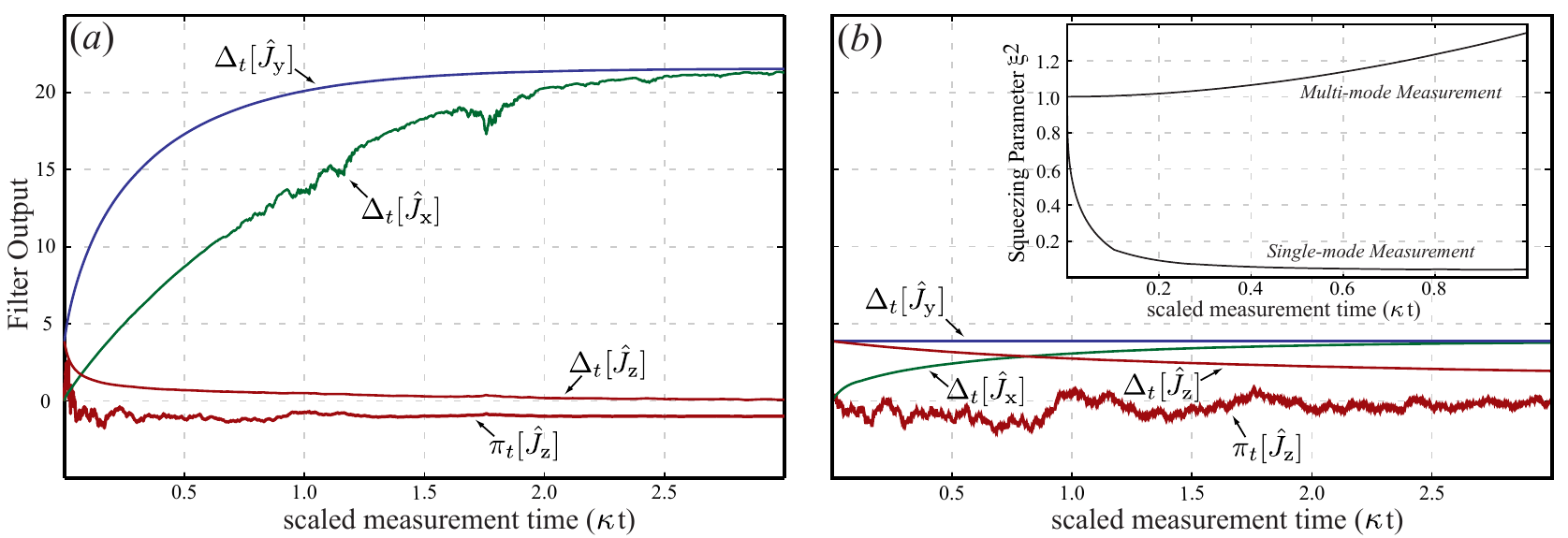} \end{center}
\vspace{-5mm}
\caption{Comparison of the single-mode $(a)$ and multi-mode $(b)$ measurement models, simulated for an ensemble of $N=60$ spin-1/2 particles with $\kappa=6$ beginning from an initial $x$-polarized coherent state.  The single-mode measurement exhibits the expected spin-squeezing and anti-squeezing; however the multi-mode measurement does not.  While there is uncertainty reduction in the collective spin $\hat{J}_\mathrm{z}$ under the multi-mode model, the squeezing parameter (inset plot) does not decrease, nor is there any anti-squeezing; the standard deviation $\Delta_t [\hat{J}_\mathrm{y}]$ remains constant.  For clarity, $\pi_t[\hat{J}_\mathrm{x}]$ and $\pi_t[\hat{J}_\mathrm{y}]$ are not plotted.  \label{Fig:CollVsSymm} }
\end{figure*}


\noindent \textit{Single-Mode (Collective) Measurement Model---} We first consider the system of indistinguishable particles interacting with a single field mode, as in Fig.\ \ref{Fig:Models}(a).  Under the interaction $\hat{H}_C =  \hbar k \hat{J}_\mathrm{z} \hat{s}_\mathrm{z}$, the atom-field system evolves according to the unitary propagator \cite{vanHandel:2005a,Chase:2009d}
\begin{align}
	d\hat{U}_t^C = \left[ \sqrt{\kappa} \hat{J}_\mathrm{z}(d\hat{S}_t^\dagger - d\hat{S}_t) - \frac{1}{2} \kappa \hat{J}_\mathrm{z}^2 dt \right] \hat{U}_t^C, \label{Eq:StochasticPropagator}
\end{align}
where $\kappa=2\pi | k(\omega_a)|^2$ is the weak-coupling interaction strength and $d\hat{S}_t$ and $d\hat{S}_t^\dagger$ are quantum Brownian motion operators satisfying the It\^o rules $d \hat{S}_t d \hat{S}_t^\dagger = dt$, $d\hat{S}_t d\hat{S}_t = d \hat{S}_t^\dagger d\hat{S}_t^\dagger = d\hat{S}_t^\dagger d\hat{S}_t = 0$ \cite{vanHandel:2005a,Chase:2009d,Hudson:1984a}.  Continuous measurement of the $\hat{s}_\mathrm{y}$ Stokes operator of the forward scattered field generates a measurement current that satisfies \cite{vanHandel:2005a,Chase:2009d}
\begin{equation}
  	dY_t^{C} = 2 \sqrt{\kappa} j_t[ \hat{J}_\mathrm{z} ] dt + d \hat{S}^\dagger_t + d \hat{S}_t .
  	\label{Eq:CollMeasurements}
\end{equation}
Equations (\ref{Eq:StochasticPropagator}-\ref{Eq:CollMeasurements}) constitute a system-observation pair, from which the conditional atomic dynamics can be inferred via the \emph{collective quantum filtering equation}
\begin{align}
	d\pi_t^C[ \hat{X} ] = & \kappa \pi_t[ \mathcal{L}^C[ \hat{J}_\mathrm{z}] \hat{X} ]dt  \label{Eq:CollectiveFilter}  \\
		& + \sqrt{ \kappa } \left( \pi_t[ \hat{J}_\mathrm{z} \hat{X} + \hat{X} \hat{J}_\mathrm{z}] - 2\pi_t[ \hat{J}_\mathrm{z}] \pi_t[ \hat{X} ] \right) dW_t^C \nonumber
\end{align}
where $dW_t^C = dY_t^C - 2 \sqrt{\kappa} \pi_t[ \hat{J}_\mathrm{z}] dt$ is a classical innovations process and $\mathcal{L}^C[ \hat{J}_z]X$ is the \emph{collective} Lindblad operator
	\begin{align}
		\mathcal{L}^C[ \hat{J}_\mathrm{z}] \hat{X} = \hat{J}_\mathrm{z} \hat{X} \hat{J}_\mathrm{z} - \frac{1}{2} \hat{J}_\mathrm{z}^2 \hat{X} - \frac{1}{2} \hat{X} \hat{J}_\mathrm{z}^2. \label{Eq:CollectiveLindblad}
	\end{align}
Note that the single-mode field model described by Eqs. (\ref{Eq:CollectiveFilter}-\ref{Eq:CollectiveLindblad}) leads to a measurement that can be formulated entirely in terms of collective angular momentum operators.
 
\noindent \textit{Multi-Mode (Symmetric) Measurement Model---} We next consider a system of particles interacting with a multi-mode field as in Fig.\ \ref{Fig:Models}(c).  Under the interaction $\hat{H}_S$, the atom-field system evolves according to
\begin{equation} \label{Eq:SymmPropagator}
	d\hat{U}_t^S =  \left[\sum_{n=1}^N  \sqrt{\kappa} \hat{j}_\mathrm{z}^{(n)} (d\hat{S}_t^{(n)\dagger} - d\hat{S}_t^{(n)})   - \frac{\kappa}{2} \hat{j}_\mathrm{z}^{(n)2} dt\right] \hat{U}_t^S
\end{equation}
where $\kappa$ is again the weak-coupling interaction strength but the $d\hat{S}_t^{(n)}$ and $d\hat{S}_t^{(n)\dagger}$ now satisfy $d \hat{S}_t^{(n)} d \hat{S}_t^{(m)\dagger} = dt \, \delta_{n,m}$ (with all other products equal to zero).  Under the condition that the multi-mode field intensity is chosen to be the same as the single mode field intensity, to achieve the same single atom-field coupling strength $\kappa$, continuous measurement of the $\hat{s}_\mathrm{y} = \sum_{n=1}^N \hat{s}_\mathrm{y}^{(n)}$ Stokes operator for the total forward scattered field generates a measurement current that satisfies \cite{Chase:2009d}
\begin{equation} \label{Eq:SymmMeasurements}
    dY_t^S = \frac{1}{\sqrt{N}}\sum_{n=1}^N 2\sqrt{\kappa} j_t[ \hat{j}_\mathrm{z}^{(n)}]dt +  d \hat{S}_t^{(n)\dagger} + d\hat{S}_t^{(n)} \vspace{-2mm}
\end{equation}
This new system-obsevations pair Eqs.\ (\ref{Eq:SymmPropagator}-\ref{Eq:SymmMeasurements}) produces the \emph{symmetric quantum filtering equation}
\begin{align}
		d\pi_t^S[\hat{X}] = &\kappa \pi_t[ \mathcal{L^S}[\hat{j}_\mathrm{z} ]\hat{X} ]dt  
			\label{Eq:SymmetricFilter} \\
		& + \sqrt{ \frac{ \kappa }{ N } } \left( \pi_t[\hat{J}_\mathrm{z} \hat{X} + \hat{X} \hat{J}_\mathrm{z}] - 2\pi_t[ \hat{J}_\mathrm{z}] \pi_t[ \hat{X} ] \right) dW_t^S, \nonumber
\end{align}
with the innovations process $dW_t^S = ( dY_t^S - 2 \sqrt{\kappa/N} \pi_t[ \hat{J}_\mathrm{z}] dt ) $ and the \textit{symmetric} Lindbladian
\begin{equation} \label{Eq:SymmetricLindblad}
		\mathcal{L}^S[\hat{j}_\mathrm{z}]\hat{X} = \sum_{n=1}^N \hat{j}_\mathrm{z}^{(n)} \hat{X} \hat{j}_\mathrm{z}^{(n)} - \frac{1}{2} \left( \hat{j}_\mathrm{z}^{(n)2} \hat{X} -  \hat{X} \hat{j}_\mathrm{z}^{(n)2}\right),
\end{equation}
which cannot be expressed using collective operators \cite{Chase:2008c}.


\noindent \textit{Squeezing and Anti-Squeezing---}  A comparison between the measurement models can be accomplished by analyzing the conditional dynamics of the collective spin operators $\hat{J}_i$.  We begin with the observed component $\hat{J}_\mathrm{z}$.   From Eqs.\ (\ref{Eq:CollectiveFilter}) and (\ref{Eq:SymmetricFilter}), the conditional $\hat{J}_\mathrm{z}$ expectation values evolve according to the filtering equations
\begin{eqnarray}
	d\pi_t^C[\hat{J}_\mathrm{z}]  & = & 2 \sqrt{ \kappa } \Delta_t^{C2}[ \hat{J}_\mathrm{z} ]
	\left(  dY_t^C - 2 \sqrt{\kappa} \pi_t[ \hat{J}_\mathrm{z}] \, dt \right) \\
	d\pi_t^S[\hat{J}_\mathrm{z}]  & = & 2 \sqrt{ \frac{\kappa }{N} } \Delta_t^{S2}[ \hat{J}_\mathrm{z} ]
	\left(  dY_t^S - 2 \sqrt{ \frac{\kappa }{N} } \pi_t[ \hat{J}_\mathrm{z}] \, dt \right). \, \, \, \, \, \, \, \, \, \, 
\end{eqnarray}
The two measurement equations are structurally similar; both are fundamentally noise-driven because $\hat{J}_\mathrm{z}$ commutes with the Lindbladians in both Eqs.\ (\ref{Eq:CollectiveFilter}) and (\ref{Eq:SymmetricFilter}).  However, the effective measurement strength in the multi-mode model is smaller by a factor of $1/\sqrt{N}$, which causes its variance $\Delta_t^{2} [\hat{J}_\mathrm{z}] = \pi_t [\hat{J}_\mathrm{z}^2] - (\pi_t [\hat{J}_\mathrm{z}])^2$ to decrease more slowly than for the single-mode measurement.  Since the evolution of the conditional expectation values depends on the variances, the multi-mode measurement not only converges more slowly, but the conditional expectations $\pi_t^C[\hat{J}_\mathrm{z}]$  and $\pi_t^S[\hat{J}_\mathrm{z}]$ do not have the same value for the two filters.

More significantly, structural differences between the models are manifest in the dynamics of operators such as $\hat{J}_\mathrm{y}^2$ because the actions of the two Lindblad terms,
\begin{equation}
	\mathcal{L}^C[\hat{J}_\mathrm{z} ]\hat{J}_\mathrm{y}^2 
		= \hat{J}_\mathrm{x}^2 - \hat{J}_\mathrm{y}^2 
	\text{   while  } 	
	\mathcal{L}^S[\hat{j}_\mathrm{z} ]\hat{J}_\mathrm{y}^2 
		=  \frac{N}{4} -\hat{J}_\mathrm{y}^2,
		\label{Eqn:JySqLindblad}
\end{equation}
are different and non-zero.  For the single-mode measurement, this term is non-zero in expectation when taken with respect to an $x$-polarized spin coherent state; it is the term responsible for anti-squeezing in the collective spin component $\hat{J}_\mathrm{y}$.  Quite remarkably, however, the multi-mode Lindblad term vanishes in expectation for an $x$-polarized state, suggesting that the multi-mode measurement generates no collective $\hat{J}_\mathrm{y}$ anti-squeezing!

 \begin{figure}[t]
\begin{center} \includegraphics{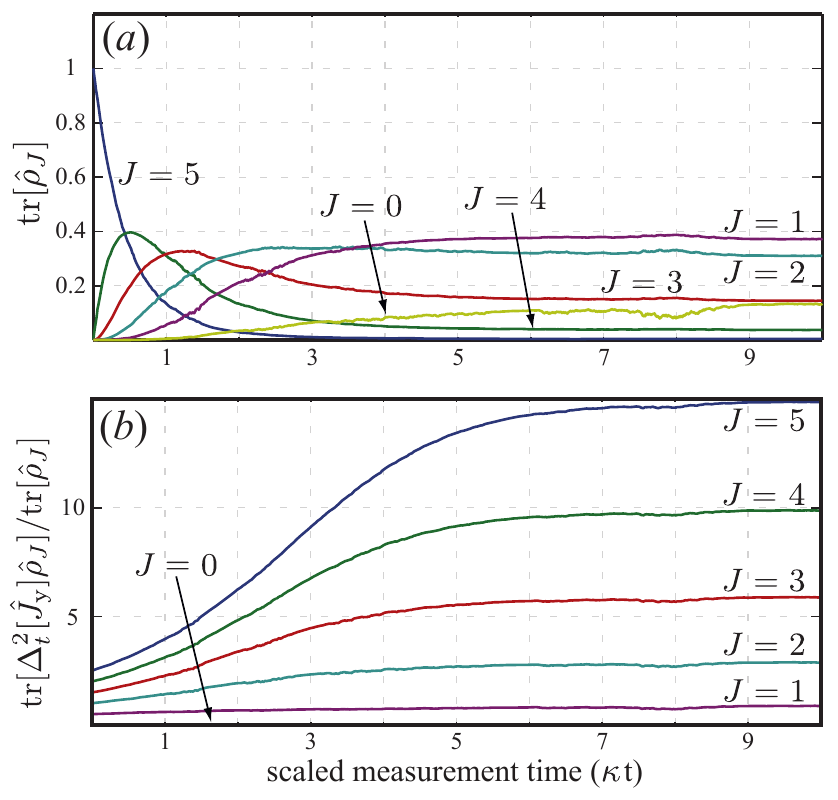} \end{center}
\vspace{-5mm}
\caption{Fig. $(a)$ plots the  traces of each total-$J$ irrep-block in the density operator Eq.\ (\ref{Eq:RhoCollective}) as it evolves from an $x$-polarized spin coherent state for $N=10$ spin-1/2 particles with $\kappa=10$. Fig. $(b)$ shows the normalized $\hat{J}_\mathrm{y}$ variances $\mathrm{tr}[ \Delta \hat{J}_\mathrm{y}^2 \hat{\rho}_J] / \mathrm{tr}[ \hat{\rho}_J]$ for each of the individual total-$J$ irrep-blocks.   \label{Fig:TracesVariances} }
\end{figure}

Figure \ref{Fig:CollVsSymm} illustrates many of these differences for an initial $x$-polarized spin coherent state subjected to the two forms of continuous measurement.  As shown by Fig.\ \ref{Fig:CollVsSymm}(a) and the inset plot, the single-mode collective model exhibits the expected spin-squeezing in $\hat{J}_\mathrm{z}$ with corresponding anti-squeezing in $\hat{J}_\mathrm{y}$.  Also as expected, the single-mode model randomly generates one of the collective $\hat{J}_\mathrm{z}$ eigenstates in the long-time limit \cite{Stockton2004}.   Figure \ref{Fig:CollVsSymm}(b) illustrates that the multi-mode symmetric model generates a different conditional evolution for $\hat{J}_\mathrm{z}$ and a slower reduction in its variance.  In fact, the $\hat{J}_\mathrm{z}$ variance decreases so slowly that the squeezing parameter $\xi_t^2 \equiv N \Delta_t^2[ \hat{J}_\mathrm{z}] / ( \pi_t^2[ \hat{J}_\mathrm{x}] +\pi_t^2[\hat{J}_\mathrm{y} ])$ never decreases below $\xi^2=1$.   Furthermore, the prediction of no anti-squeezing from Eq.\ (\ref{Eqn:JySqLindblad}) is confirmed by the simulation--- the variance $\Delta^2_t[\hat{J}_\mathrm{y}]=N/4$ is a constant of the measurement dynamics.   The multi-mode measurement evidently does not produce a conventional spin-squeezed state \cite{Kitagawa1993,Wineland:1994a}.

\vspace{2mm}

\noindent\textit{Conditional Quantum State Dynamics}---   The only structural difference in the two filters, Eqs.\ (\ref{Eq:CollectiveFilter}) and (\ref{Eq:SymmetricFilter}), resides in the form of their Lindbladians.   Dynamics that can be expressed entirely in terms of collective spin operators, such as the single-mode Lindbladian Eq.\ (\ref{Eq:CollectiveLindblad}), preserve states that are invariant under the permutation of particle labels.   Provided that the initial state is permutation invariant, the dynamics are then confined to an ($N+1$)-dimensional sub-Hilbert space corresponding to the maximum $\hat{J}^2$ eigenvalue $J_\mathrm{max}=N/2$, the so-called \textit{symmetric group}  \cite{Stockton2003a}.   Since the multi-mode Lindbladian Eq.\ (\ref{Eq:SymmetricLindblad}) is not generated by collective spin operators, it can be shown to couple the different total-$J$ irreducible representations (irreps) of the rotation group \cite{Chase:2008c}.  The dynamics are not restricted to the symmetric group, but rather the $O(N^2)$-dimensional sub-Hilbert space that is invariant across degenerate copies of the different total-$J$ irreps.  Such states are called \textit{generalized collective states} and are described by density operators of the form \cite{Chase:2008c}
\begin{equation} \label{Eq:RhoCollective}
	\hat{\rho}_C = \bigoplus_J \hat{\rho}_J =  \sum_{J,M,M'} \rho_{J,M,M'} \ketbra{J,M}{J,M'}, 
\end{equation}
i.e, the direct sum over irrep-blocks $\hat{\rho}_J$ corresponding to different total spin $J=0,1,2,\ldots, N/2$.

The form of the symmetric Lindbladian and the resulting coupling between total-$J$ irrep-blocks accounts for the decoherence due to ignoring the ``which-mode" information discussed in Fig \ref{Fig:Models}.  Evolution under the multi-mode measurement does not preserve pure states (even in the case of perfect detection efficiency) because the states become mixed over total-$J$ irrep-blocks.  This behavior is demonstrated by the simulated measurement in Fig.\ \ref{Fig:TracesVariances}(a), which plots the traces of the individual irrep-blocks $\hat{\rho}_J$ as the measurement evolves in time.  Analyzing the behavior of the individual irrep-blocks also helps to explain the absence of collective anti-squeezing.   Owing to the structure of Eq.~(\ref{Eq:RhoCollective}), the total variance in $\hat{J}_\mathrm{y}$ is the sum over the variances of the individual $J$-irrep-blocks. Fig.\ \ref{Fig:TracesVariances}(b) reveals that these individual irrep variances exhibit anti-squeezing, exactly as though each respective block was subject to its own $\hat{J}_\mathrm{z}$ measurement.    The total $\hat{J}_\mathrm{y}$ variance does not increase because the symmetric Lindbladian transfers population to irrep-blocks $\hat{\rho}_J$ with decreasing total spin as the measurement proceeds.

 \begin{figure}[t]
\begin{center} \includegraphics{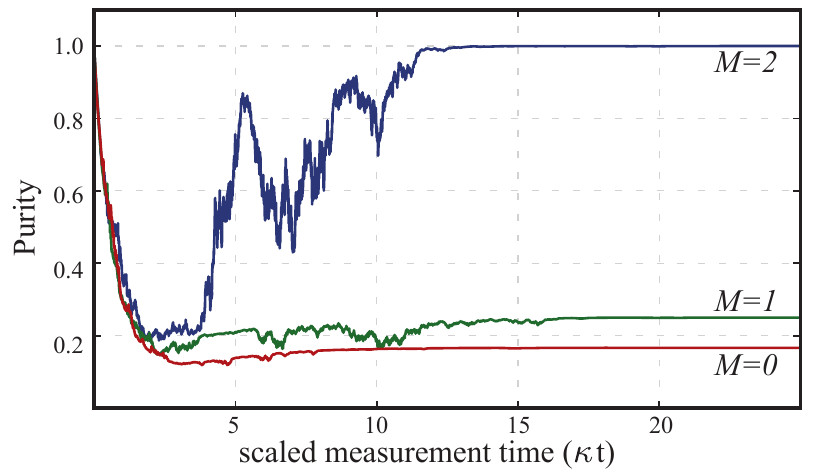} \end{center}
\vspace{-6mm}
\caption{Purity comparison of the steady states $M = \{2,1,0\}$ for an ensemble of 4 particles with $\kappa=25$, with respective values \{1, 1/4, 1/6\} according to $1/\alpha_4^M$.   $N=4$ was chosen for reasons of clarity. \label{Fig:Purity}}
\end{figure} 

A further distinction between the models lies in the long-time behavior of the filtering equations.   Like the single-mode measurement, the steady states of the multi-mode measurement exhibit zero collective variance $\Delta_\infty^2[\hat{J}_\mathrm{z}] \to 0$ and thus a well-defined angular momentum $J_\mathrm{z} = \hbar M$.   The steady states are therefore conveniently labeled by the quantum number $M$,
\begin{equation} \label{Eq:SteadyStates}
		\hat{\rho}_M^\mathrm{ss} = \frac{1}{\alpha^M_N} \sum_{J \geq M} d_N^{J} {\ketbra{J,M}{J,M}},
\end{equation}
where $d_N^J =N!(2J+1) / (N/2-J)!(N/2 + J + 1)!$ is the number of degenerate irreps with total spin $J$ and $\alpha_N^{M}$ is the cumulative sum $ \alpha_N^{M} = \sum_{J \geq M} d^{J}_N$.  However, in stark contrast to the single-mode model, which always converges to a pure $\hat{J}_\mathrm{z}$ eigenstate \cite{Stockton2004}, $\hat{\rho}_M$ is a heavily mixed state with purity $1/\alpha_N^M$. Fig. \ref{Fig:Purity} shows the time-evolution of the purity as the multi-mode measurement converges to different steady states.   The only way for the measurement to generate a final pure state is in the unlikely case that $M = J_\mathrm{max}$.  For $N \gg 1$, typical realizations will produce states with vanishing purity, since $d_N^J$ grows exponentially in $N$ for all irreps other than $J=J_\mathrm{max}$. 

\vspace{2mm}

\noindent\textit{Conclusion---} We have found that the common practice of focusing a multi-mode probe field onto a single detector does not produce a true collective spin measurement for dilute atomic samples.  Previous theoretical treatments based on symmetric states fail to account for the decoherence that results from ignoring ``which-particle'' information that is available in principle for spatially-resolvable particles. While typical experiments will generally lie somewhere between the limiting cases of Figs.\ \ref{Fig:Models}(a) and \ref{Fig:Models}(c), one would still expect measurement models based entirely on symmetric states and collective spin operators \cite{Kitagawa1993,Stockton2003a,Stockton2004} to overestimate significantly the expected degree of squeezing and anti-squeezing.  Such models are likely inadequate to describe dispersive measurements performed by coupling a large spin ensemble to a probe laser field in free space.
 
We thank Rob Cook for helpful discussions.  This work was supported by the NSF under grants PHY-0639994 and PHY-0652877.  Please visit http://qmc.phys.unm.edu/ to download the simulation code used to generate our results as well as all data files used to generate the figures in this paper.

\vspace{-2mm}

\bibliography{Paper}

\end{document}